\documentclass[espf]{aa}

\usepackage{color} 

\usepackage{graphicx}
\usepackage{amsmath}
\usepackage[psamsfonts]{euscript}
\usepackage[psamsfonts]{amssymb}
\usepackage{amsmath}
\usepackage{graphicx}
\usepackage{natbib}
\usepackage{epsfig}

\titlerunning{}

\def\kms{km~s$^{-1}$}

\begin{document}
\title{High resolution spectroscopy for Cepheids distance
determination\thanks{Based on observations made with ESO telescopes
at the Silla Paranal Observatory under programme IDs 072.D-0419 and
073.D-0136}}


\subtitle{V. Impact of the cross-correlation method on the p-factor
and the $\gamma$-velocities}
\titlerunning{High resolution spectroscopy for Cepheids distance
determination V.}
\authorrunning{N. Nardetto and collaborators}

\author{N. Nardetto \inst{1}, W. Gieren\inst{1}, P.
Kervella\inst{2}, P. Fouqu\'e\inst{3}, J. Storm\inst{4}, G.
Pietrzynski \inst{1,5}, D. Mourard\inst{6}, D. Queloz\inst{7}}

\institute{Departamento de Astronom\'ia, Universidad de
Concepci\'on, Casilla 160-C, Concepci\'on \and Observatoire de
Paris-Meudon, LESIA, UMR 8109, 5 Place Jules Janssen, F-92195 Meudon
Cedex, France \and Observatoire Midi-Pyr\'en\'ees, Laboratoire
d'Astrophysique, UMR 5572, Universit\'e Paul Sabatier - Toulouse 3,
14 avenue Edouart Belin, 31400 Toulouse, France \and
Astrophysikalisches Institut Postdam, An der Sternwarte 16, 14482,
Postdam, Germany \and Warsaw University Observatory, AL. Ujazdowskie
4, 00-478, Warsaw, Poland \and OCA/CNRS/UNS, Dpt. Fizeau, UMR6525,
Avenue Copernic, F-06130 Grasse, France \and Observatoire de
Gen\`eve, Universit\'e de Gen\`eve, 51 Ch. des Maillettes, 1290
Sauverny, Switzerland}


\date{Received ... ; accepted ...}

\abstract{The cross correlation method (hereafter \emph{CC}) is
widely used to derive the radial velocity curve of Cepheids when the
signal to noise of the spectra is low. However, if it is used with
the ``wrong'' projection factor, it might introduce some biases in
the Baade-Wesselink (hereafter BW) methods of determining the
distance of Cepheids. In addition, it might affect the average value
of the radial velocity curve (or $\gamma$-velocity) important for
Galactic structure studies.} {We aim to derive a period-projection
factor relation (hereafter \emph{Pp}) appropriate to be used
together with the CC method. Moreover, we investigate whether the CC
method can explain the misunderstood previous calculation of the
``K-term'' of Cepheids.} {We observed eight galactic Cepheids with
the HARPS\thanks{High Accuracy Radial velocity Planetary Search
project developed by the European Southern Observatory}
spectrograph. For each star, we derive an interpolated CC radial
velocity curve using the HARPS pipeline. The amplitudes of these
curves are used to determine the correction to be applied to the
semi-theoretical projection factor derived in Nardetto et al.
(2007). Their average value (or $\gamma$-velocity) are also compared
to the center-of-mass velocities derived in Nardetto et al. (2008).}
{The correction in amplitudes allows us to derive a new \emph{Pp}
relation: $p = [-0.08 \pm 0.05] \log P + [1.31 \pm 0.06]$. We also
find a negligible wavelength dependence (over the optical range) of
the \emph{Pp} relation. We finally show that the $\gamma$-velocity
derived from the CC method is systematically blue-shifted by about
$1.0 \pm 0.2$\kms compared to the center-of-mass velocity of the
star. An additional blue-shift of $1.0$\kms is thus needed to
totally explain the previous calculation of the ``K-term'' of
Cepheids (around 2\kms).} {The new \emph{Pp} relation we derived is
a solid tool for the distance scale calibration, and especially to
derive the distance of LMC Cepheids with the infrared surface
brightness technique. Further studies should be devoted to determine
the impact of the signal to noise ratio, the spectral resolution,
and the metallicity on the \emph{Pp} relation.}


\maketitle

\section{Introduction}\label{s_Introduction}

The Baade-Wesselink (hereafter \emph{BW}) method of determining the
distance of Cepheids was recently used to calibrate the
period-luminosity (\emph{PL}) of Galactic Cepheids (Fouqu\'e et al.
2007). The basic principle of this method is to compare the linear
and angular size variation of a pulsating star in order to derive
its distance through a simple division. The angular diameter is
either derived by interferometry (for e.g. Kervella et al. 2004,
Davis et al. 2008) or using the infrared surface brightness
(hereafter IRSB) relation (Gieren et al. 1998, 2005a). However, when
determining the linear radius variation of the Cepheid by
spectroscopy, one has to use a conversion projection factor from
radial to pulsation velocity. This quantity has been studied using
hydrodynamic calculations by Sabbey et al. (1996), and more recently
Nardetto et al. (2004, 2007).

Following the work of Burki et al. (1982), we showed in Nardetto et
al. (2006, hereafter Paper I) that the first moment of the spectral
line is the only method which is independent of the spectral line
width (average value and variation) and the rotation velocity of the
star. The centroid radial velocity ($RV_{\mathrm c}$), or the first
moment of the spectral line profile, is defined as
\begin{equation} \label{Eq_CDG}
 RV_{\mathrm c} = \frac{\int_{\rm line} \lambda S(\lambda) d\lambda}{\int_{\rm line} S(\lambda) d\lambda}
\end{equation}
We thus used this definition of the radial velocity in paper two of
this series (Nardetto et al. 2007, hereafter Paper II), to derive a
semi-theoretical period-projection factor (hereafter \emph{Pp})
relation based on spectroscopic measurements with the HARPS high
resolution spectrograph. This relation was derived from the specific
\ion{Fe}{I} 4896.439 \AA\, spectral line which has a relatively low
depth for all stars at all pulsation phase (around 8\% of the
continuum). It was shown that such low-depth is suitable to reduce
the uncertainty on the projection factor due to the velocity
gradient which takes place between the photosphere (corresponding to
angular diameter measurements) and the line-forming region
(corresponding to the radius estimation from spectroscopic
measurements).

In the cross-correlation method (hereafter CC method) a mask
(composed of hundreds or thousands) of spectral lines is convolved
to the observed spectrum. The resulting average profile is then
fitted by a Gaussian. In such a method, there is first a mix of
different spectral lines forming at different levels (more or less
sensitive to a velocity gradient). Second, the resulting velocity
can be dependent of the abundances or effective temperature (through
the line width), or the rotation of the stars. Third, in paper III
of this series (Nardetto et al. 2008) we derived calibrated
center-of-mass velocities of the stars of our HARPS sample. By
comparing these so-called $\gamma$-velocities with the ones found in
the literature (generally based on the CC method) and in particular
in the Galactic Cepheid Database (Fernie et al. 1995), we obtained
an average correction of $1.8 \pm 0.2$ \kms. This result shows that
the ``K-term'' of Cepheids stems from an intrinsic property of
Cepheids. But, it does show also that the cross-correlation might
introduce a bias (up to a few kilometers per second) on the average
value of the radial velocity curve.

After a careful definition of the projection factor (Sect.
\ref{s_Definition}), we apply the cross-correlation method to the
Cepheids of our HARPS sample (sect. \ref{s_data}), in order to
derive a period-projection factor relation appropriated to the CC
method (Sect. \ref{s_Ppcc}). As the HARPS pipeline also provides the
cross-correlated radial velocities for each spectral order, we take
the opportunity to study the wavelength dependence of the projection
factor law (Sect. \ref{s_SpecAna}). And finally, we quantify the
impact of the CC method on the $\gamma$-velocities (Sect.
\ref{s_Kterm}).

\section{Definition of the ``CC projection
factor''}\label{s_Definition}

In this section, we recall some results obtained in paper~II and we
define the projection factor well suitable for the cross-correlation
method. In paper II, we defined the projection factor as:
\begin{equation} \label{Eq_pf}
p=\frac{\Delta V_{\mathrm{p}}^{\mathrm{o}}}{\Delta RV_\mathrm{c}}
\end{equation} where $\Delta V_{\mathrm{p}}^{\mathrm{o}}$ is the amplitude of the
pulsation velocity curve associated to the photosphere of the star.
$\Delta RV_{\mathrm{c}}$ is the amplitude of the radial velocity
curve obtained from the first moment of the spectral line. Because
of the atmospheric velocity gradient, $\Delta RV_{\mathrm{c}}$
depends on the spectral line considered. Using a selection of 17
spectral lines, we thus derived an interpolated relation between
$\Delta RV_{\mathrm{c}}$ and $D$, where $D$ is the line depth
corresponding to the minimum radius of the star :
\begin{equation} \label{Eq_gradient}
\Delta RV_{\mathrm{c}} = a_0 D + b_0
\end{equation}

This relation was then used to quantify the correction
($f_{\mathrm{grad}}$) to be applied on the projection factor due to
the velocity gradient (see Eq.~3 of paper II). The
\ion{Fe}{I}~4896.439~\AA\ spectral line (which forms close to the
photosphere) was found to provide the lowest correction. The
amplitude of the radial velocity curve corresponding to the
\ion{Fe}{I} 4896.439~\AA\ spectral line was finally used (see
$f_{\mathrm{grad}}$ in Tab. 5 of paper II) to derive the
semi-theoretical \emph{Pp} relation. It is defined (Eq.
\ref{Eq_gradient}) as $\Delta RV_{\mathrm{c}}[\mathrm{4896}] = a_0
D_{\mathrm{ 4896}} + b_0$, where $a_0$ and $b_0$ are indicated in
Tab. 3 of Paper II. $D_{\mathrm{4896}}$ is derived from the
interpolation of the line depth curve at the particular phase
corresponding to the minimum radius of the star (i.e. when
$RV_\mathrm{c}$ corrected from the $\gamma$-velocity $\simeq 0$).
$D_{\mathrm{4896}}$ and $\Delta RV_{\mathrm{c}}[\mathrm{4896}]$ are
given in Tab. \ref{Tab1} of this paper.

\begin{table}[h]
\begin{center}
\caption[]{The cepheids studied listed with increasing period. The
depth of the \ion{Fe}{I} 4896.439 \AA\ spectral line and the
corresponding value of the amplitude of the radial velocity curve
(see Eq. \ref{Eq_gradient}) are given.}\label{Tab1}
\begin{tabular}{lcccccc}
\hline \hline \noalign{\smallskip}

Cepheid        &    $P^{\mathrm{\tiny (a)}}$   &     $D_{\mathrm{ 4896}}$     &   $\Delta RV_{\mathrm{c}} [\mathrm{4896}]$   \\
            &    [days]                     &          [\%]                &           [\kms ]             \\
\hline
R TrA       &    $3.38925$                &         6      &   $  28.6   _{\pm   0.5    }$   \\
S Cru       &    $4.68976$                &         6      &   $  33.5   _{\pm   0.5    }$   \\
Y Sgr       &    $5.77338$                &         5      &   $  34.3   _{\pm   0.5    }$   \\
$\beta$ Dor &    $9.84262$                &        10      &   $  31.7   _{\pm   0.5    }$   \\
$\zeta$ Gem &    $10.14960$               &        14      &   $  25.5   _{\pm   0.5    }$   \\
RZ Vel      &   $20.40020$               &         4      &   $  47.6   _{\pm   0.5    }$   \\
$\ell$~Car  &    $35.55134$               &        13      &   $  32.8   _{\pm   0.5    }$   \\
RS Pup      &    $41.51500$               &         4      &   $  42.4   _{\pm   0.5    }$   \\

\hline \noalign{\smallskip}
\end{tabular}
\end{center}
\begin{list}{}{}

\item[$^{\mathrm{a}}$] The corresponding Julian dates
($T_{\mathrm{o}}$) can be found in Paper~II.

\end{list}
\end{table}

The projection factor suitable to the cross-correlation method
(hereafter $p_{\mathrm cc}$) is then simply :
\begin{equation} \label{Eq_Ppcc}
p_{\mathrm cc} = p \frac{\Delta RV_{\mathrm{c}}[\mathrm
4896]}{\Delta RV_\mathrm{cc}} = p f_{\mathrm{cc}}
\end{equation}
where ${\Delta RV_\mathrm{cc}}$ is the amplitude of the radial
velocity curve obtained with the cross-correlation method, and
$f_{\mathrm{cc}}$ the correcting factor to be applied. Our
definition of $p_{\mathrm cc}$ is independent of the
$\gamma$-velocities.

\begin{figure*}[htbp]
\begin{center}
\begin{tabular}{cc}
\resizebox{1.0\hsize}{!}{\includegraphics[clip=true]{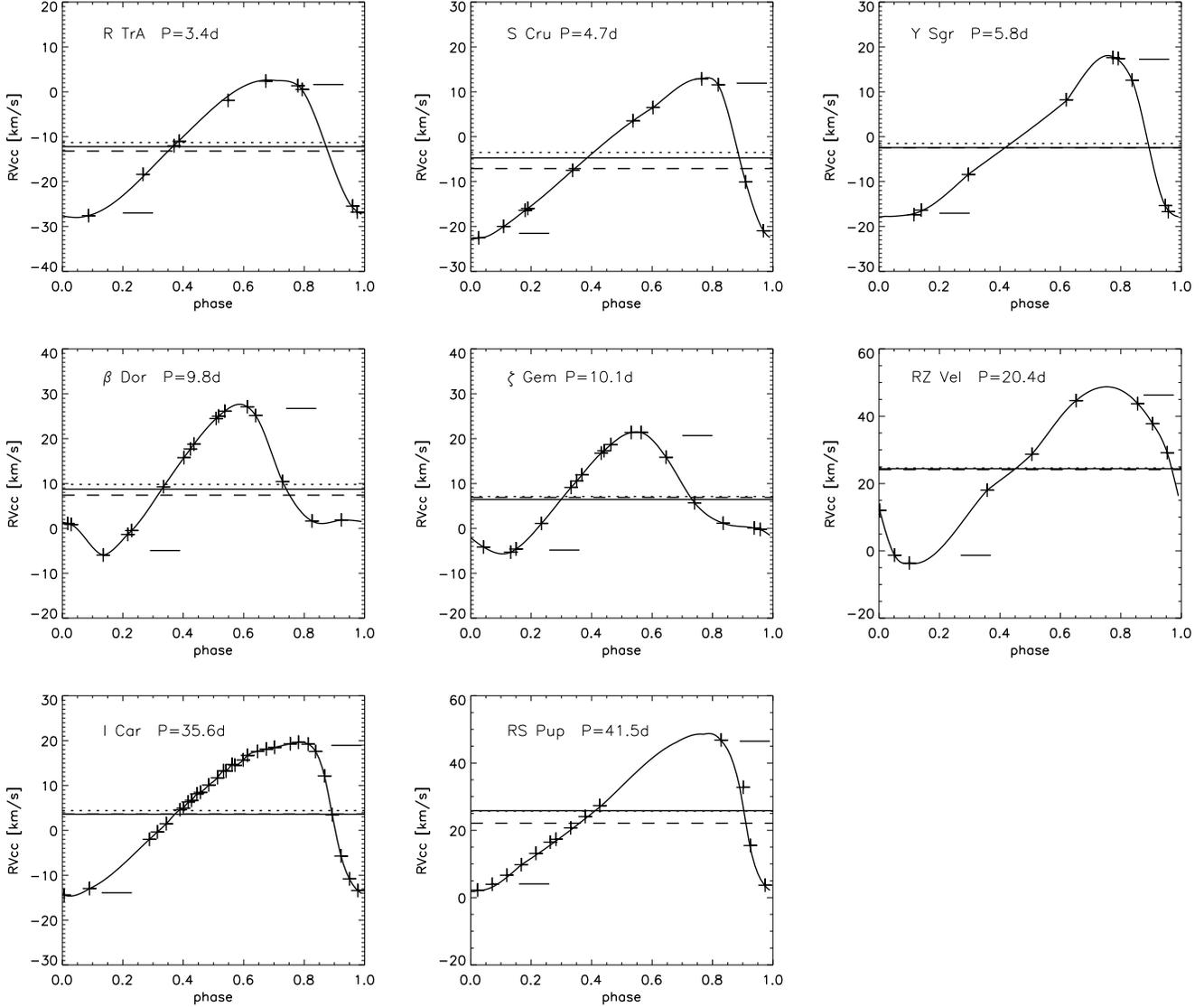}} &
\end{tabular}
\end{center}
\caption{Interpolated radial velocity curves based on the
cross-correlation method are presented of each Cepheid in our
sample. Uncertainty are too small to be seen (around 0.5 \kms). The
horizontal lines near {\it extrema} give an indication about $\Delta
RV_{\mathrm{c}}[\mathrm{4896}]$. The short horizontal lines are the
$\gamma$-velocities (see Sect. \ref{s_Kterm}) corresponding to the
CC method (solid line), the center-of-mass velocity of paper III
(dotted line) and from Fernie et al. (1995, dashed line).}
\label{f1}
\end{figure*}

\section{The CC method applied to HARPS
observations}\label{s_data}

We consider eight Cepheids which have been observed with the HARPS
spectrometer ($R=120000$): \object{R~Tra}, \object{S~Cru},
\object{Y~Sgr}, \object{$\beta$~Dor}, \object{$\zeta$~Gem},
\object{RZ~Vel}, \object{$\ell$~Car}, \object{RS~Pup}. Information
about observations (number of measurements, pulsation phases) can be
found in Paper~I.

We apply the HARPS pipeline to our data in order to calculate the
cross-correlated radial velocities (Baranne et al. 1996, Pepe et al.
2002). The basic principle of the CC method is to build a mask, made
of zero and non-zero value-zones, where the non-zero zones
correspond to the theoretical positions and widths of thousands of
metallic spectral lines at zero velocity, carefully selected from a
synthetic spectrum of a G2 star. A relative weight is considered for
each spectral line according to its depth (derived directly from
observations of a G2 type star). An average spectral line profile is
finally constructed by shifting the mask as a function of the
Doppler velocity. The corresponding radial velocity is derived
applying a classical $\chi^{2}$ minimization algorithm between the
observed line profile and a Gaussian function. The whole profil is
considered in the fitting procedure, not only the line core. The
average value of the fitted Gaussian corresponds to the
cross-correlated radial velocity (hereafter $RV_{\mathrm cc}$). The
HARPS instrument has 72 spectral orders. The pipeline provides
$RV_{\mathrm cc}$ averaged over the 72 spectral orders, or
independently for each order. We first use the averaged values and
the corresponding uncertainties.

\begin{table*}
\begin{center}
\caption[]{The quantities used to derive the CC projection factor
($p_{\mathrm{cc}}$) are listed. We compare also the
$\gamma$-velocities ($V_{\gamma}${\tiny [CC]}) derived from the CC
radial velocity curves with previous results.}\label{Tab2}
\begin{tabular}{lcccc|ccc}
\hline \hline \noalign{\smallskip}

Name          & $p$ {\tiny (a)}                  &     $\Delta RV_{\mathrm{cc}}$     &       $f_{\mathrm{cc}}$            &    $p_{\mathrm{cc}}$          &     $V_{\gamma}${\tiny [GCD](b)}     &   $V_{\gamma}${\tiny [N08](c)}     &   $V_{\gamma}$ {\tiny [CC]}             \\
              &                                   &         [\kms]                    &                                   &                               &          [\kms]                        &        [\kms]                       &       [\kms]                   \\
\hline
R~TrA   &   $   1.35    _{\pm   0.03    }$  &   $   30.6    _{\pm   0.6 }$  &   $   0.93    _{\pm   0.03    }$  &   $   1.25    _{\pm   0.07    }$  &   $   -13.2   _{\pm   2.0 }$  &   $   -11.3   _{\pm   0.3 }$  &   $   -12.2   _{\pm   0.6 }$  \\
S~Cru   &   $   1.34    _{\pm   0.03    }$  &   $   35.9    _{\pm   0.6 }$  &   $   0.93    _{\pm   0.03    }$  &   $   1.25    _{\pm   0.07    }$  &   $   -7.1    _{\pm   2.0 }$  &   $   -3.5    _{\pm   0.4 }$  &   $   -4.7    _{\pm   0.6 }$  \\
Y~Sgr   &   $   1.33    _{\pm   0.03    }$  &   $   36.0    _{\pm   0.5 }$  &   $   0.95    _{\pm   0.03    }$  &   $   1.26    _{\pm   0.05    }$  &   $   -2.5    _{\pm   2.0 }$  &   $   -1.5    _{\pm   0.2 }$  &   $   -2.4    _{\pm   0.5 }$  \\
$\beta$~Dor &   $   1.32    _{\pm   0.02    }$  &   $   33.5    _{\pm   0.2 }$  &   $   0.94    _{\pm   0.02    }$  &   $   1.24    _{\pm   0.05    }$  &   $   7.4 _{\pm   2.0 }$  &   $   9.8 _{\pm   0.1 }$  &   $   8.7 _{\pm   0.2 }$  \\
$\zeta$~Gem &   $   1.31    _{\pm   0.02    }$  &   $   27.2    _{\pm   0.2 }$  &   $   0.94    _{\pm   0.03    }$  &   $   1.23    _{\pm   0.05    }$  &   $   6.9 _{\pm   2.0 }$  &   $   7.1 _{\pm   0.1 }$  &   $   6.4 _{\pm   0.2 }$  \\
RZ~Vel  &   $   1.30    _{\pm   0.02    }$  &   $   52.5    _{\pm   0.7 }$  &   $   0.91    _{\pm   0.02    }$  &   $   1.18    _{\pm   0.05    }$  &   $   24.1    _{\pm   2.0 }$  &   $   24.6    _{\pm   0.4 }$  &   $   24.4    _{\pm   0.7 }$  \\
$\ell$~Car  &   $   1.28    _{\pm   0.02    }$  &   $   34.3    _{\pm   0.2 }$  &   $   0.96    _{\pm   0.02    }$  &   $   1.22    _{\pm   0.04    }$  &   $   3.6 _{\pm   2.0 }$  &   $   4.4 _{\pm   0.1 }$  &   $   3.6 _{\pm   0.2 }$  \\
RS~Pup  &   $   1.27    _{\pm   0.02    }$  &   $   46.9    _{\pm   0.5 }$  &   $   0.90    _{\pm   0.02    }$  &   $   1.16    _{\pm   0.04    }$  &   $   22.1    _{\pm   2.0 }$  &   $   25.7    _{\pm   0.2 }$  &   $   25.8    _{\pm   0.5 }$  \\

\hline \noalign{\smallskip}
\end{tabular}
\end{center}
\begin{list}{}{}
\item[$^{\mathrm{a}}$] The projection factor as derived in paper II.
\item[$^{\mathrm{b}}$] The $\gamma$-velocities derived from the
Galactic Cepheid Database (Fernie et al. 1995).
\item[$^{\mathrm{c}}$] The $\gamma$-velocities or calibrated
center-of-mass velocities of the stars from paper III.
\end{list}
\end{table*}

The $RV_{\mathrm cc}$ curves are then carefully interpolated using a
periodic cubic spline function. This function is calculated either
directly on the observational points or using arbitrary pivot
points. In the latter case, a classical minimization process between
observations and the interpolated curve is used to optimize the
position of the pivot points (M\'erand et al. 2005). For Y~Sgr and
RS~Pup, pivot points are used due to an inadequate phase coverage.
When the phase coverage is good (which is the case for all other
stars), the two methods are equivalent (Fig. \ref{f1}). From these
curves we are finally able to calculate $\Delta RV_{\mathrm{cc}}$
(Tab. \ref{Tab2}). The statistical uncertainty on $\Delta
RV_{\mathrm{cc}}$ is set as the average value of the uncertainty
obtained for all measurements over a pulsation cycle of the star.

\section{A \emph{Pp} relation dedicated to the CC method}\label{s_Ppcc}

\begin{figure}[htbp]
\begin{center}
\resizebox{1.0\hsize}{!}{\includegraphics[clip=true]{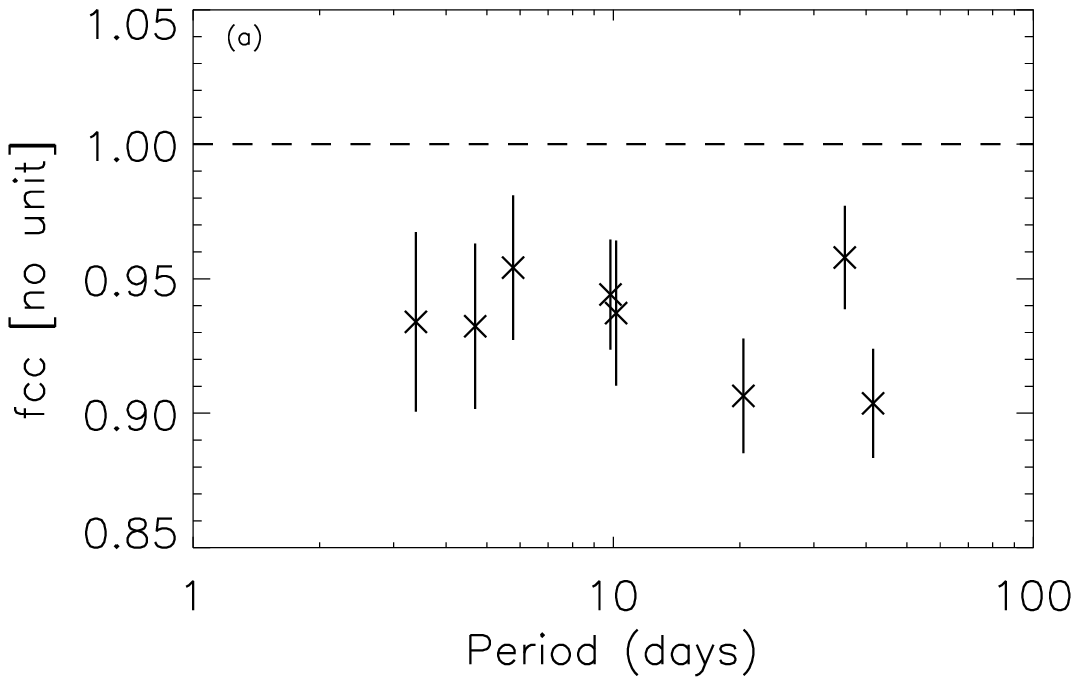}}
\resizebox{1.0\hsize}{!}{\includegraphics[clip=true]{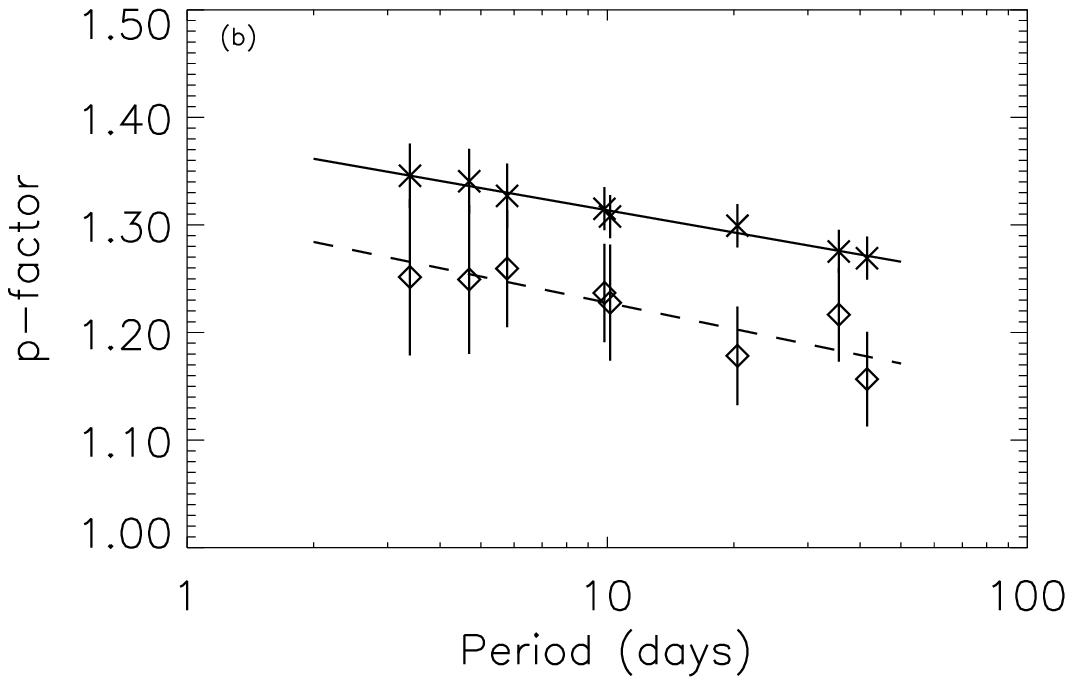}}
\end{center}
\caption{(a) The correction factor $f_{\mathrm{cc}}$ induced on the
projection factor by the cross-correlation method is shown as a
function of the logarithm of the period of the star. (b) The
period-projection factor ($p$) relation from paper II (crosses and
solid line) and the corrected relation suitable for the
cross-correlation method $p_{\mathrm{cc}}=p f_{\mathrm cc}$
(diamonds and dashed line).} \label{f2f3}
\end{figure}

From $\Delta RV_{\mathrm{c}} [\mathrm{4896}]$ and ${\Delta
RV_\mathrm{cc}}$ obtained for all stars we derive the correcting
factor $f_{\mathrm{cc}}$ using Eq. \ref{Eq_pf}. The result is
plotted as a function of the period in Fig. \ref{f2f3}a. No
particular trend is found. However, the $f_{\mathrm{cc}}$ correction
factors are clearly statistically dispersed around a mean value of
$0.93 \pm 0.02$.

Following our definition ($p_{\mathrm{cc}} = p f_{\mathrm{cc}}$),
the corrected projection factors suitable for the CC method are
given in Tab. \ref{Tab2}. The relation between the period and
$p_{\mathrm{cc}}$ remains clear according to the statistical
uncertainties:

\begin{equation} \label{Eq_Ppcc_result}
p_{\mathrm{cc}} = p f_{\mathrm{cc}} = [-0.08 \pm 0.05] \log P +
[1.31 \pm 0.06].
\end{equation}

The corresponding reduced $\chi^2$ is 1.2. We refer to this relation
in the following using $Pp_{\mathrm{cc}}$. We recall that the
\emph{Pp} relation we found in Paper II dedicated to the FeI
\emph{4896} spectral line was : $p = [-0.064 \pm 0.020] \log P +
[1.376 \pm 0.023]$. These two relations are shown in Fig.
\ref{f2f3}b. The impact of the cross-correlation method on the
zero-point of the \emph{Pp} is thus significant, while the slope
increases only slightly (in absolute value) from $-0.064$ to
$-0.08$.

We have several possible explanations for these results. The
cross-correlation induces actually two biases:

\begin{enumerate}

\item the cross-correlated radial velocities are
derived using a Gaussian fit, making the result sensitive both to
the spectral line width (i.e. the effective temperature and
abundances) and the rotation velocity projected on the line of
sight. These two quantities, independently, and even more the
combination of both, are not really expected to vary linearly with
the logarithm of the period. It might explain why no clear linear
relation is found between $f_{\mathrm{cc}}$ and the period of the
star. However, the mean value of the correction factors (around
$0.93 \pm 0.02$) have a non negligible impact on the zero-point of
the $Pp$ relation which decreases from $1.376$ to $1.31$ (5\%).

\item the cross-correlation method implies a mix of different
spectral lines forming at different levels. In the $Pp$ relation,
the only quantity sensitive to the line depth is $f_{\mathrm{grad}}$
(as defined in Paper II) which compares the amplitude of the
pulsation velocity corresponding to the line-forming region, and the
photosphere, respectively. It is thus an estimate of the velocity
gradient within the pulsating atmosphere of the star. The $Pp$
relation was derived in paper~II for the \emph{4896} spectral line
which forms very close to the photosphere ($D=8\%$), while the
cross-correlated radial velocity is a mix of thousands of spectral
lines forming at different levels, with an average depth of around
$D \simeq 25\%$. The cross-correlation method is thus more sensitive
to the velocity gradient (because the average line depth is large),
which may explain the increase (in absolute value) of the slope from
$-0.064$ to $-0.08$. Moreover, in paper II we provided a very rough
estimate of the $Pp$ relation associated to the cross-correlation
method considering {\it only} the impact of the velocity gradient
(which means discarding the bias related to the Gaussian fit). We
found $p = [-0.075 \pm 0.031] \log P + [1.366 \pm 0.036]$ (see Sect.
7 of Paper II). The slope we find here ($-0.08$) is consistent with
this previous rough estimate of $-0.075$.
\end{enumerate}

These results are important to take into account when deriving the
distance of Galactic or LMC/SMC Cepheids using the cross-correlation
method. We emphasize that our $Pp_{\mathrm{cc}}$ is consistent with
the result by M\'erand et al. (2005), who found $p=1.27$ for
$\delta$~Cep ($P=5.36$).

\section{Wavelength dependence of the projection factor}\label{s_SpecAna}

\begin{figure}[htbp]
\begin{center}
\resizebox{1.0\hsize}{!}{\includegraphics[clip=true]{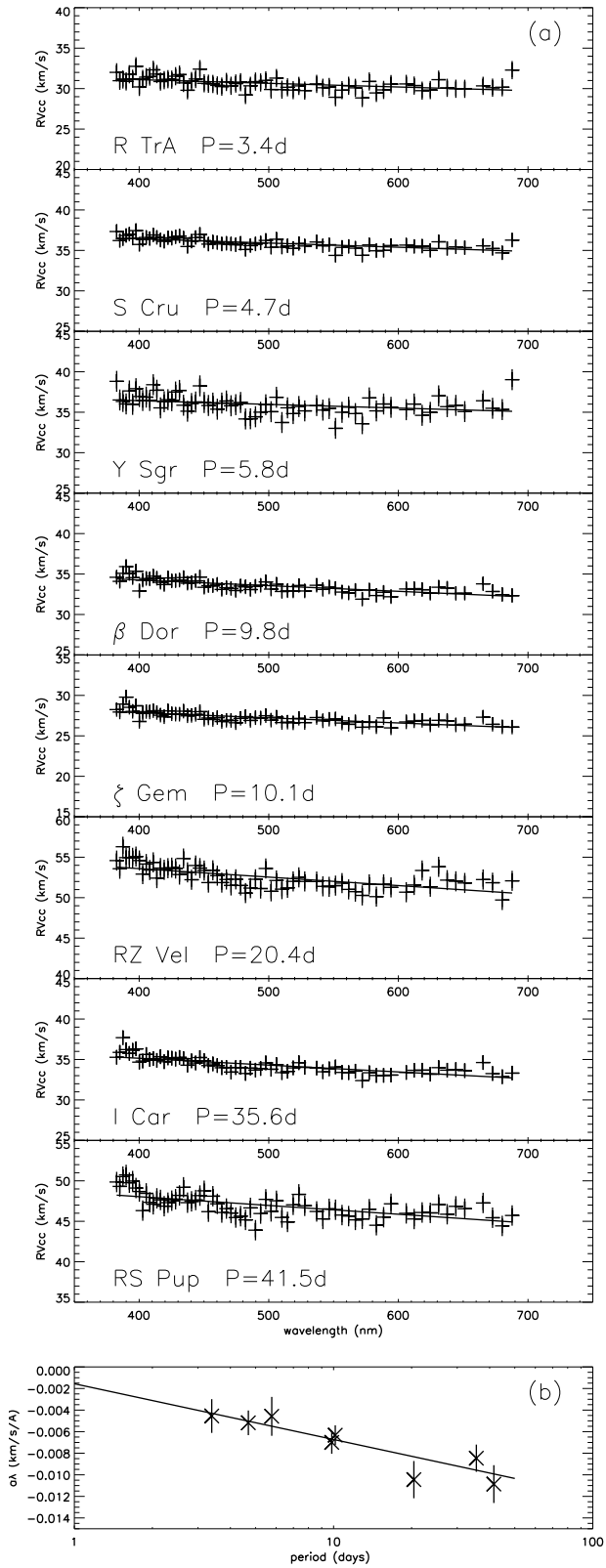}}
\end{center}
\caption{(a) Wavelength dependency of the amplitude of the
cross-correlated radial velocity curves for each star in our sample.
The corresponding linear relation are defined as : $\Delta
RV_{\mathrm{cc}} = a_{\mathrm{\lambda}} \lambda +
b_{\mathrm{\lambda}} $ (b) The corresponding slopes
($a_{\mathrm{\lambda}}$) as a function of the period.}
\label{Fig_f5}
\end{figure}

With the data at hand, we check for a possible dependence of the
projection factor on the wavelength range used for the
cross-correlation radial velocity measurement. For each order, we
derive the cross-correlated interpolated radial velocity curves, and
then the corresponding amplitudes ${\Delta RV_\mathrm{cc}}
\mathrm{(\lambda)}$. Orders 59, 68 and 72 are not considered due to
instrumental characteristics and/or unrealistic results. For all
stars, ${\Delta RV_\mathrm{cc}} \mathrm{(\lambda)}$ is plotted as a
function of the wavelength, defined as the orders' average values
(Fig. \ref{Fig_f5}a). We find linear relations between these two
quantities:

\begin{equation} \label{Eq_RVcc_lambda}
{\Delta RV_\mathrm{cc}} \mathrm{(\lambda)} = a_{\mathrm{\lambda}}
\lambda + b_{\mathrm{\lambda}},
\end{equation}
where $a_{\mathrm{\lambda}}$ and $b_{\mathrm{\lambda}}$ are listed
in Tab. \ref{Tab3}. For consistency with the previous section the
${\Delta RV_\mathrm{cc}} \mathrm{(\lambda)}$ quantities have been
slightly shifted in velocity in such a way that :
\begin{equation} \label{Eq_RVcc_lambda2}
{\Delta RV_\mathrm{cc}} = a_{\mathrm{\lambda}} 502.2 \mathrm{nm} +
b_{\mathrm{\lambda}},
\end{equation}
where $\Delta RV_{\mathrm{cc}}$ is derived from Tab. \ref{Tab2} and
$502.2$nm is the wavelength averaged over all orders.

We also find a relation between $a_{\mathrm{\lambda}}$ and the
logarithm of the period of the star :
\begin{equation} \label{Eq_a_lambda}
a_{\lambda} = [-0.005\pm0.001] \log P - [0.002 \pm 0.001]
\end{equation}

From these results we can make two comments. First, the amplitude of
the cross-correlated radial velocity curves decreases with the
wavelength. From hydrodynamical modelling, we know that the spectral
lines form over a larger part of the atmosphere in the infrared
compared to optical (Sasselov et al. 1990). This effect might a key
to understand our result: the more extended the line forming regions
are, the lower is the amplitude of the radial velocity curves.
Second, this effect is larger for long-period Cepheids as compared
to short-period Cepheids. A reason might be that the mean radius,
the size of the line-forming regions and the velocity gradient
increase with the logarithm of the period.

In order to quantify the wavelength dependency of the
$Pp_{\mathrm{cc}}$ relation, we define two correction factors
($f_{\mathrm{\lambda=400nm}}=\frac{\Delta RV_{\mathrm{cc}}
\mathrm{(\lambda=400nm)}}{\Delta RV_{\mathrm{cc}}}$ and
$f_{\mathrm{\lambda=700nm}}=\frac{\Delta RV_{\mathrm{cc}}
\mathrm{(\lambda=700nm)}}{\Delta RV_{\mathrm{cc}}}$). We find the
following correcting relation as a function of the logarithm of the
period:

\begin{equation} \label{Eq_RVcc_flambdaB}
f_{\mathrm{\lambda=400nm}}=[-0.01 \pm 0.01] \log P + [0.99 \pm
0.01],
\end{equation}

and

\begin{equation} \label{Eq_RVcc_flambdaR}
f_{\mathrm{\lambda=700nm}}=[0.02 \pm 0.01] \log P + [1.02 \pm 0.01],
\end{equation}

\begin{figure}[htbp]
\begin{center}
\resizebox{1.0\hsize}{!}{\includegraphics[clip=true]{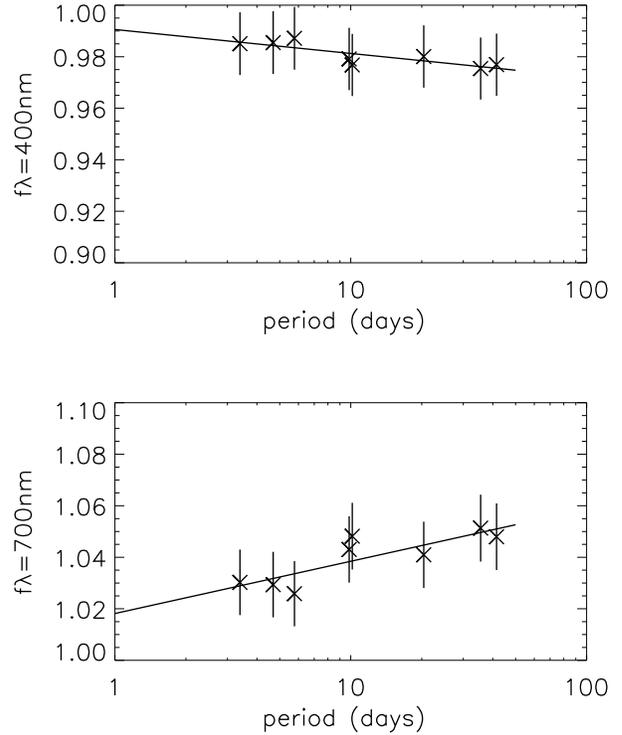}}
\end{center}
\caption{Corrections to apply to the $Pp_{\mathrm{cc}}$ relation
(Eq. \ref{Eq_Ppcc_result}) in the blue ($\lambda=400$nm) and in the
red ($\lambda=700$nm).} \label{Fig_f6}
\end{figure}

\begin{table}
\begin{center}
\caption[]{Coefficients of the linear relations between the
amplitude of the radial velocity curve and the
wavelength.}\label{Tab3}
\begin{tabular}{lcc}
\hline \hline \noalign{\smallskip}

Name          &            $a_{\lambda}$                  &     $b_{\lambda}$                   \\
              &                                   &            \\
\hline
R~TrA   &   $   -0.005  _{\pm   0.002   }$  &   $   32.93   _{\pm   0.80    }$  \\
S~Cru   &   $   -0.005  _{\pm   0.001   }$  &   $   38.50   _{\pm   0.58    }$  \\
Y~Sgr   &   $   -0.005  _{\pm   0.002   }$  &   $   38.26   _{\pm   0.92    }$  \\
$\beta$~Dor &   $   -0.007  _{\pm   0.001   }$  &   $   37.06   _{\pm   0.54    }$  \\
$\zeta$~Gem &   $   -0.006  _{\pm   0.001   }$  &   $   30.39   _{\pm   0.46    }$  \\
RZ~Vel  &   $   -0.010  _{\pm   0.002   }$  &   $   57.76   _{\pm   0.88    }$  \\
$\ell$~Car  &   $   -0.008  _{\pm   0.001   }$  &   $   38.54   _{\pm   0.64    }$  \\
RS~Pup  &   $   -0.011  _{\pm   0.002   }$  &   $   52.39   _{\pm   0.89    }$  \\

\hline \noalign{\smallskip}
\end{tabular}
\end{center}
\end{table}

The reduced $\chi^2$ are respectively 0.3 and 1.3. These relations
are shown in Fig. \ref{Fig_f6}. We find that such corrections are
currently irrelevant given our statistical uncertainties on the
$Pp_{\mathrm{cc}}$ relation (Eq. \ref{Eq_Ppcc_result}).

\section{The CC $\gamma$-velocity and the K-term of Cepheids}\label{s_Kterm}

Interestingly, for each Cepheid in our sample, we found in paper III
a linear relation between the $\gamma$-velocities (derived using the
first moment method) of the various spectral lines and their
corresponding $\gamma$-asymmetries. Using these linear relations, we
provided a physical reference to derive the center-of-mass
$\gamma$-velocity of the stars ($V_{\gamma}${\tiny [N08]}): it
should be zero when the $\gamma$-asymmetry is zero. These values are
consistent with a axisymmetric rotation model of the Galaxy.
Conversely, previous measurements of the $\gamma$-velocities found
in the literature (for e.g. Fernie et al. 1995 : the Galactic
Cepheid Database, hereafter $V_{\gamma}${\tiny [GCD]}) were based on
the cross-correlation method, and by using generally only few
measurements over the pulsation cycle. These results led to an
apparent ``fall'' of Galactic Cepheids towards the Sun (compared to
an axisymmetric rotation model of the Milky Way) with a mean
velocity of about 2 km/s. This residual velocity shift has been
dubbed the ``K-term'', and was first estimated by Joy (1939) to be
-3.8 km/s.

\begin{figure}[htbp]
\begin{center}
\resizebox{1.0\hsize}{!}{\includegraphics[clip=true]{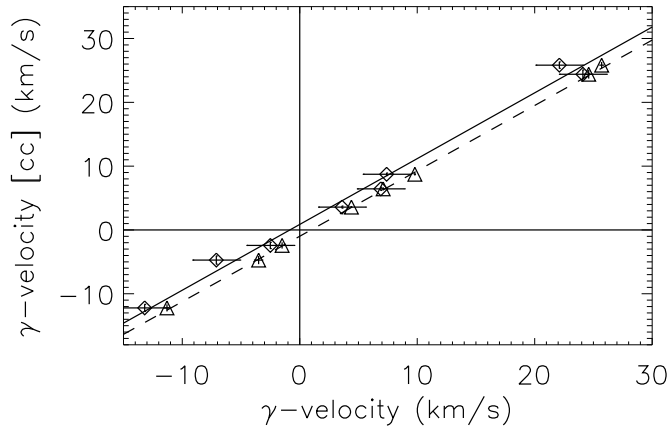}}
\end{center}
\caption{$V_{\gamma}$ {\tiny [CC]} as a function of
$V_{\gamma}${\tiny [GCD]} (diamond) and $V_{\gamma}${\tiny [N08]}
(triangles). The solid and dashed line are the corresponding linear
interpolation respectively.} \label{f4}
\end{figure}

We aim to understand why such a 2km/s mistake was done before. An
hypothesis is that the cross-correlation method is biased by the
dynamical structure of the atmosphere of Cepheids. To verify this
hypothesis, we have the unique opportunity to compare quantitatively
and in a consistent way $V_{\gamma}${\tiny [N08]},
$V_{\gamma}${\tiny [GCD]} and the $\gamma$-velocities derived from
our HARPS cross-correlated radial velocity curves (hereafter
$V_{\gamma}${\tiny [CC]}). The comparison is done by plotting
$V_{\gamma}${\tiny [CC]} as a function of $V_{\gamma}${\tiny [GCD]}
and $V_{\gamma}${\tiny [N08]} respectively (Fig. \ref{f4}). Data are
all presented in Tab. \ref{Tab2} and the resulting linear relations
are respectively :

\begin{equation} \label{Eq_Vg1}
V_{\gamma} {\mathrm{\tiny [CC]}} = [1.03 \pm 0.06]
V_{\gamma}{\mathrm{\tiny [GCD]}} + [0.86 \pm 0.78] ,
\end{equation}

 and

\begin{equation} \label{Eq_Vg1}
V_{\gamma} {\mathrm{\tiny [CC]}} = [1.02 \pm 0.02]
V_{\gamma}{\mathrm{\tiny [N08]}} - [0.99 \pm 0.17],
\end{equation}

The reduce $\chi^2$ are respectively $3.0$ and $3.8$.

Several conclusions must be pointed out. The slope of these
relations are similar and close to one, which basically means that
there is no particular trend of the $\gamma$-velocity with the
period of the star, or at least, it remains negligible here. As in
paper III, we find a systematic difference of $0.86 + 0.99 \simeq
1.8$\kms between $V_{\gamma}${\tiny [N08]} and $V_{\gamma}${\tiny
[GCD]}, which is consistent with the K-term of Cepheids. However,
the $\gamma$-velocities we derive in this paper using the
cross-correlation method are systematically lower by ($0.9 \pm 0.8$)
\kms than the ones found in the literature ($V_{\gamma}${\tiny
[GCD]}), and they are systematically larger by $1.0 \pm 0.2$ \kms
than the calibrated center-of-mass velocities ($V_{\gamma}${\tiny
[N08]}). As a consequence, the cross-correlation method alone cannot
explain alone the K-term. The CC method is sensitive in such a way
to the dynamical structure of Cepheids' atmosphere, that it is
responsible for $\simeq$50\% of the K-term. Something else is
requested to explain the presence of such offsets in previous
determinations of the gamma-velocity. It could be related, for
instance, to the quality of observations in the past (Joy et al.
1939) or to the different methods used to derive the
$\gamma$-velocity (Pont et al. 1994).

\section{Conclusions}

By comparing the amplitude of our cross-correlated radial velocity
curves with previous results based on the first moment method (paper
II), we derived a new \emph{Pp} relation applicable to radial
velocities measured by the cross-correlation method. This relation
is crucial for the distance scale calibration, and in particular to
derive the distances of LMC and SMC Cepheids (Gieren et al. 2005a;
Gieren et al. 2009, in preparation). We find also a slight
dependence of the \emph{Pp} relation on the wavelength. Considering
our current uncertainties this effect is negligible, but it might
become significant in the near future. The next steps are certainly
to test the impact of the signal to noise ratio, the spectral
resolution and the metallicity on the projection factor. The latter
point will require a large sample of Cepheids with well-measured
metallicities. These studies (including this work) are fully part of
the international ``Araucaria Project'' whose purpose is to provide
an improved local calibration of the extragalactic distance scale
out to distances of a few Megaparsecs (Gieren et al. 2005b).
Moreover, the fact that the cross-correlation method over-estimates
the amplitude of the radial velocity curve and under-estimates the
$\gamma$-velocity (compared to the calibrated values presented in
paper III) might have some implications for other kinds of pulsating
stars, for e.g. in asteroseismology.

Moreover, we show in paper III that the K-term of Cepheids vanished
if one considers carefully the dynamical structure of Cepheids
atmosphere. From the results presented in this paper, we can state
that the cross-correlation method might not be totally responsible
for the K-term found in the previous studies (only 50\% seems to be
a consequence of the cross-correlation method). There seems to be
another contribution whose nature should be investigated.

\begin{acknowledgements}
Based on observations collected at La Silla observatory, Chile, in
the framework of European Southern Observatory's programs 072.D-0419
and 073.D-0136. This research has made use of the SIMBAD and VIZIER
databases at CDS, Strasbourg (France). NN and WG acknowledge
financial support from the FONDAP Center of Astrophysics 15010003,
and the BASAL Center of Astrophysics CATA. NN acknowledges the
Geneva team for his support in using the HARPS pipeline.
\end{acknowledgements}


\begin{table*}
\begin{center}
\caption[]{HARPS cross-correlated radial velocities (part 1)
\label{Tab_online1}}
\begin{tabular}{|cc|cc|cc|cc|cc|}
\hline \hline \noalign{\smallskip}

         \multicolumn{2}{|c|}{R TrA}    &  \multicolumn{2}{|c|}{S Cru}    &  \multicolumn{2}{|c|}{Y Sgr}       &   \multicolumn{2}{|c|}{$\beta$~Dor} & \multicolumn{2}{|c|}{$\zeta$~Gem}    \\
phase  &  $RV_{\mathrm{cc}} $   [km/s]&   phase  &  $RV_{\mathrm{cc}}$ [km/s]& phase  &  $RV_{\mathrm{cc}}$   [km/s]& phase  &  $RV_{\mathrm{cc}}$ [km/s]& phase  &  $RV_{\mathrm{cc}}$  [km/s] \\

\hline 

\noalign{\smallskip}

0.09    &   $   -27.59  \pm 0.65    $   &   0.03    &   $   -22.53  \pm 0.62    $   &   0.12    &   $   -17.37  \pm 0.65    $   &   0.02    &   $   1.08    \pm 0.18    $   &   0.04    &   $   -4.16   \pm 0.33    $   \\
0.27    &   $   -18.40  \pm 0.65    $   &   0.11    &   $   -19.98  \pm 0.66    $   &   0.14    &   $   -16.35  \pm 0.44    $   &   0.03    &   $   0.86    \pm 0.17    $   &   0.13    &   $   -5.29   \pm 0.20    $   \\
0.37    &   $   -12.05  \pm 0.43    $   &   0.18    &   $   -16.40  \pm 0.42    $   &   0.30    &   $   -8.43   \pm 0.61    $   &   0.14    &   $   -5.94   \pm 0.24    $   &   0.15    &   $   -4.59   \pm 0.17    $   \\
0.39    &   $   -11.01  \pm 0.56    $   &   0.19    &   $   -15.99  \pm 0.59    $   &   0.62    &   $   8.22    \pm 0.55    $   &   0.22    &   $   -1.35   \pm 0.14    $   &   0.23    &   $   1.10    \pm 0.17    $   \\
0.55    &   $   -1.91   \pm 0.48    $   &   0.34    &   $   -7.50   \pm 0.49    $   &   0.77    &   $   17.62   \pm 0.46    $   &   0.23    &   $   -0.48   \pm 0.15    $   &   0.33    &   $   9.12    \pm 0.18    $   \\
0.67    &   $   2.39    \pm 0.70    $   &   0.54    &   $   3.55    \pm 0.62    $   &   0.79    &   $   17.40   \pm 0.40    $   &   0.33    &   $   9.28    \pm 0.21    $   &   0.35    &   $   10.54   \pm 0.28    $   \\
0.78    &   $   1.36    \pm 0.42    $   &   0.60    &   $   6.51    \pm 0.50    $   &   0.84    &   $   12.60   \pm 0.43    $   &   0.40    &   $   15.77   \pm 0.21    $   &   0.37    &   $   11.94   \pm 0.20    $   \\
0.79    &   $   0.53    \pm 0.61    $   &   0.76    &   $   12.89   \pm 0.77    $   &   0.95    &   $   -15.33  \pm 0.39    $   &   0.42    &   $   17.70   \pm 0.16    $   &   0.43    &   $   16.76   \pm 0.30    $   \\
0.96    &   $   -25.47  \pm 0.46    $   &   0.82    &   $   11.56   \pm 0.52    $   &   0.96    &   $   -16.71  \pm 0.45    $   &   0.44    &   $   18.76   \pm 0.23    $   &   0.44    &   $   17.31   \pm 0.18    $   \\
0.98    &   $   -26.80  \pm 0.63    $   &   0.91    &   $   -10.10  \pm 1.40    $   &       &   $               $   &   0.51    &   $   24.50   \pm 0.20    $   &   0.46    &   $   18.69   \pm 0.33    $   \\
    &   $               $   &   0.97    &   $   -20.96  \pm 0.51    $   &       &   $               $   &   0.52    &   $   24.99   \pm 0.19    $   &   0.53    &   $   21.36   \pm 0.27    $   \\
    &   $               $   &       &   $               $   &       &   $               $   &   0.54    &   $   26.14   \pm 0.28    $   &   0.56    &   $   21.37   \pm 0.17    $   \\
    &   $               $   &       &   $               $   &       &   $               $   &   0.61    &   $   27.10   \pm 0.27    $   &   0.65    &   $   15.84   \pm 0.39    $   \\
    &   $               $   &       &   $               $   &       &   $               $   &   0.64    &   $   25.14   \pm 0.25    $   &   0.74    &   $   5.69    \pm 0.51    $   \\
    &   $               $   &       &   $               $   &       &   $               $   &   0.73    &   $   10.42   \pm 0.18    $   &   0.84    &   $   1.15    \pm 0.17    $   \\
    &   $               $   &       &   $               $   &       &   $               $   &   0.83    &   $   1.64    \pm 0.14    $   &   0.94    &   $   0.08    \pm 0.17    $   \\
    &   $               $   &       &   $               $   &       &   $               $   &   0.92    &   $   1.88    \pm 0.16    $   &   0.96    &   $   -0.25   \pm 0.22    $   \\

  \hline \hline \noalign{\smallskip}
\end{tabular}
\end{center}
\end{table*}

\begin{table*}
\begin{center}
\caption[]{HARPS cross-correlated radial velocities (part 2)
\label{Tab_online2}}
\begin{tabular}{|cc|cc|cc|}
\hline \hline \noalign{\smallskip}

         \multicolumn{2}{|c|}{RZ Vel}    &  \multicolumn{2}{|c|}{$\ell$~Car}    & \multicolumn{2}{|c|}{RS Pup}    \\
phase  &  $RV_{\mathrm{cc}} $   [km/s]&   phase  & $RV_{\mathrm{cc}}$ [km/s]& phase  &  $RV_{\mathrm{cc}}$   [km/s]\\

\hline 

\noalign{\smallskip}

0.00    &   $   12.01   \pm 1.50    $   &   0.01    &   $   -14.44  \pm 0.14    $   &   0.02    &   $   2.23    \pm 0.41    $   \\
0.05    &   $   -1.31   \pm 0.67    $   &   0.09    &   $   -12.98  \pm 0.16    $   &   0.07    &   $   3.97    \pm 0.49    $   \\
0.10    &   $   -3.62   \pm 0.81    $   &   0.29    &   $   -2.01   \pm 0.25    $   &   0.12    &   $   6.66    \pm 0.41    $   \\
0.36    &   $   18.03   \pm 0.42    $   &   0.31    &   $   -0.38   \pm 0.15    $   &   0.17    &   $   9.79    \pm 0.48    $   \\
0.51    &   $   28.71   \pm 0.32    $   &   0.34    &   $   1.45    \pm 0.18    $   &   0.22    &   $   13.14   \pm 0.36    $   \\
0.65    &   $   44.64   \pm 0.51    $   &   0.39    &   $   4.58    \pm 0.14    $   &   0.26    &   $   16.48   \pm 0.37    $   \\
0.86    &   $   43.72   \pm 0.77    $   &   0.40    &   $   4.94    \pm 0.13    $   &   0.28    &   $   17.34   \pm 0.37    $   \\
0.90    &   $   37.78   \pm 0.60    $   &   0.42    &   $   6.32    \pm 0.19    $   &   0.33    &   $   20.71   \pm 0.45    $   \\
0.95    &   $   29.14   \pm 0.55    $   &   0.43    &   $   6.67    \pm 0.18    $   &   0.38    &   $   24.06   \pm 0.54    $   \\
    &   $               $   &   0.45    &   $   8.12    \pm 0.13    $   &   0.43    &   $   27.32   \pm 0.38    $   \\
    &   $               $   &   0.46    &   $   8.45    \pm 0.18    $   &   0.83    &   $   46.81   \pm 0.59    $   \\
    &   $               $   &   0.48    &   $   10.08   \pm 0.22    $   &   0.90    &   $   32.78   \pm 0.65    $   \\
    &   $               $   &   0.51    &   $   11.69   \pm 0.18    $   &   0.93    &   $   15.48   \pm 0.87    $   \\
    &   $               $   &   0.53    &   $   13.27   \pm 0.28    $   &   0.97    &   $   3.70    \pm 0.60    $   \\
    &   $               $   &   0.54    &   $   13.17   \pm 0.20    $   &       &   $               $   \\
    &   $               $   &   0.56    &   $   14.68   \pm 0.17    $   &       &   $               $   \\
    &   $               $   &   0.57    &   $   14.51   \pm 0.20    $   &       &   $               $   \\
    &   $               $   &   0.60    &   $   15.67   \pm 0.19    $   &       &   $               $   \\
    &   $               $   &   0.61    &   $   16.69   \pm 0.14    $   &       &   $               $   \\
    &   $               $   &   0.65    &   $   17.59   \pm 0.16    $   &       &   $               $   \\
    &   $               $   &   0.67    &   $   18.10   \pm 0.13    $   &       &   $               $   \\
    &   $               $   &   0.70    &   $   18.48   \pm 0.13    $   &       &   $               $   \\
    &   $               $   &   0.75    &   $   19.28   \pm 0.13    $   &       &   $               $   \\
    &   $               $   &   0.78    &   $   19.61   \pm 0.13    $   &       &   $               $   \\
    &   $               $   &   0.81    &   $   19.20   \pm 0.11    $   &       &   $               $   \\
    &   $               $   &   0.84    &   $   17.57   \pm 0.15    $   &       &   $               $   \\
    &   $               $   &   0.87    &   $   12.12   \pm 0.15    $   &       &   $               $   \\
    &   $               $   &   0.89    &   $   3.45    \pm 0.17    $   &       &   $               $   \\
    &   $               $   &   0.92    &   $   -5.73   \pm 0.13    $   &       &   $               $   \\
    &   $               $   &   0.95    &   $   -10.82  \pm 0.12    $   &       &   $               $   \\
    &   $               $   &   0.98    &   $   -13.41  \pm 0.17    $   &       &   $               $   \\

  \hline \hline \noalign{\smallskip}
\end{tabular}
\end{center}
\end{table*}

\end{document}